\documentclass{article}

\usepackage{arxiv}

\usepackage[utf8]{inputenc} % allow utf-8 input
\usepackage[T1]{fontenc}    % use 8-bit T1 fonts
\usepackage{hyperref}       % hyperlinks
\usepackage{url}            % simple URL typesetting
\usepackage{booktabs}       % professional-quality tables
\usepackage{amsfonts}       % blackboard math symbols
\usepackage{nicefrac}       % compact symbols for 1/2, etc.
\usepackage{microtype}      % microtypography
\usepackage{lipsum}		% Can be removed after putting your text content
\usepackage{graphicx}
\usepackage{natbib}
\usepackage{doi}
\usepackage{lipsum}

\title{``You tell me'': A Dataset of GPT-4-Based Behaviour Change Support Conversations}

\author{ \href{https://orcid.org/0000-0002-4736-2565}{\includegraphics[scale=0.06]{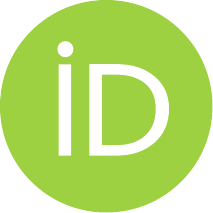}\hspace{1mm}Selina Meyer}\\
	Regensburg University\\
	Regensburg, Germany\\
	\texttt{selina.meyer@ur.de} \\
	\And
	\href{https://orcid.org/0000-0002-5791-0641}{\includegraphics[scale=0.06]{orcid.pdf}\hspace{1mm}David Elsweiler} \\
	Regensburg University\\
	Regensburg, Germany\\
	\texttt{david.elsweiler@co.uk} \\
}

\renewcommand{\shorttitle}{Meyer and Elsweiler}

\hypersetup{
pdftitle={``You tell me'': A Dataset of GPT-4-Based Behaviour Change Support Conversations},
pdfauthor={Selina Meyer, David Elsweiler},
}

\begin{document}
\maketitle

\begin{abstract}
Conversational agents are increasingly used to address emotional needs on top of information needs. One use case of increasing interest are counselling-style mental health and behaviour change interventions, with large language model (LLM)-based approaches becoming more popular. Research in this context so far has been largely system-focused, foregoing the aspect of user behaviour and the impact this can have on LLM-generated texts. To address this issue, we share a dataset containing text-based user interactions related to behaviour change with two GPT-4-based conversational agents collected in a preregistered user study. This dataset includes conversation data, user language analysis, perception measures, and user feedback for LLM-generated turns, and can offer valuable insights to inform the design of such systems based on real interactions.
\end{abstract}

% keywords can be removed

\begin{figure}[htbp]
    \centering
    \includegraphics[width=0.5\columnwidth]{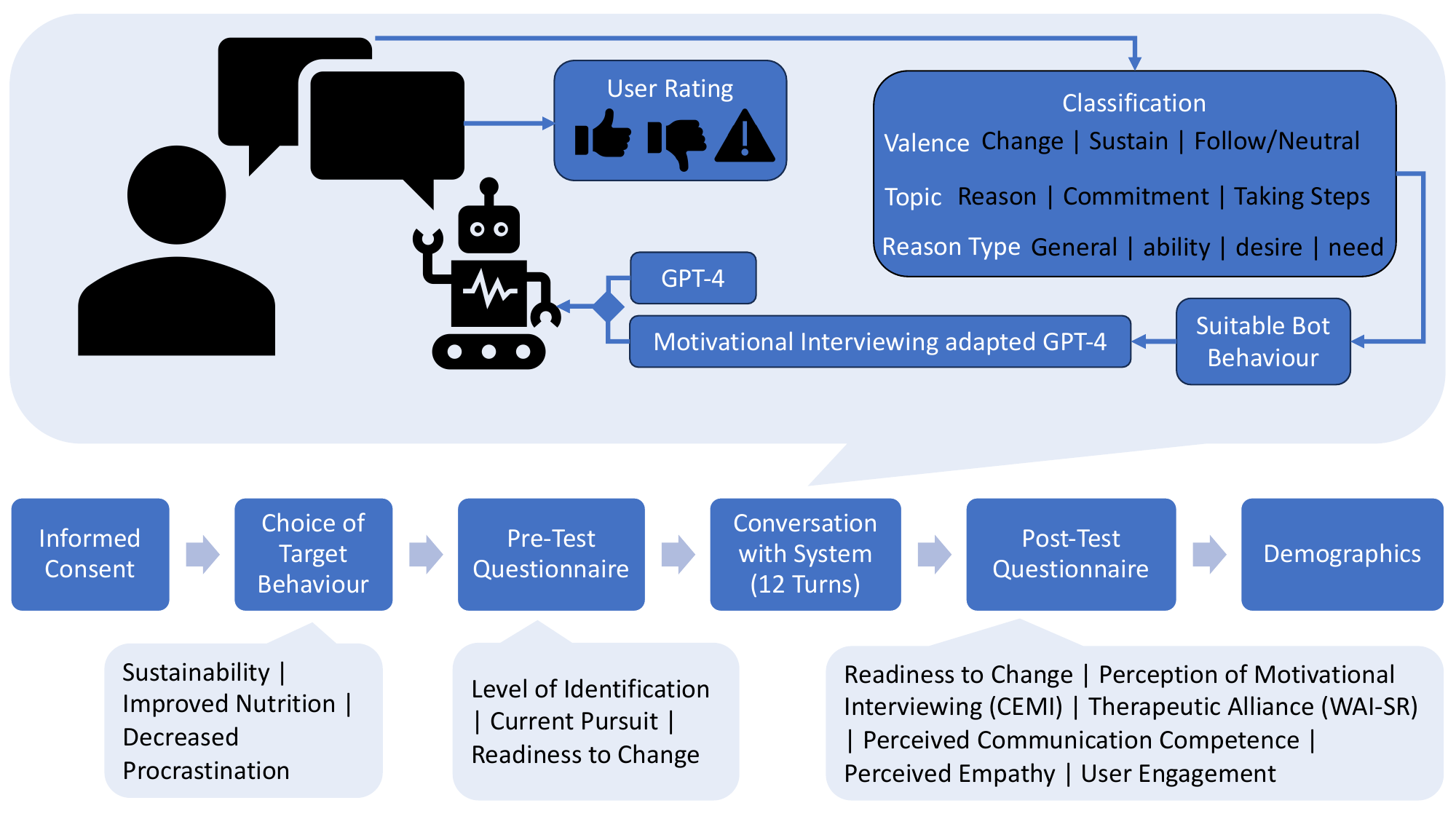}
    \caption{Each user interacts with one of two systems, where one system is prompted to adhere to Motivational Interviewing principles. Users interact with the systems for 12 turns. User turns are classified with respect to their implications regarding motivation for behaviour change. Each GPT-generated bot turn is rated as helpful, unhelpful, or harmful  by the user, with an optional rating explanation.}
    \label{fig:overview}
\end{figure}

\section{Introduction}
Chat interactions with conversational agents (CAs) are often studied in task-oriented domains, such as customer service \citep{elsholz2019exploring}, e-commerce \citep{papenmeier2021dataset, papenmeier2022mhm}, or cooking \citep{barko2020conversational, frummet2022can, nouri2020step}. In these scenarios, the CA and the user work together to solve a clear information need. However, information systems and other forms of online interaction are not only utilised to address information needs, but are also popular for leisure, or purely hedonistic purposes \citep{elsweiler2011understanding} or to address emotional or social needs \citep{ruthven2019making}. Past research has highlighted the importance of meeting emotional needs when designing conversational agents, especially in sensitive contexts \citep{lopatovska2023designing, rapp2021human, meyer2022m}.

Consequently, there has been an increased interest in social influence dialogue systems, systems that automate behaviour and health interventions and are focused on addressing emotional needs more than information needs in recent years \citep{chawla-etal-2023-social}. One conversational strategy that has been explored in this context is Motivational Interviewing (MI), a therapy approach aimed at increasing a person's motivation to change through supportive, non-confrontational conversation \citep{miller2012motivational}. So far, conversational systems in this context have predominantly employed rule- or retrieval-based strategies \citep{xu2022survey, he2022can, park2019designing, schulman2011intelligent, olafsson2019coerced, boustani2021development, samrose2022mia}. However, rule-based conversational agents are often restrictive and fail to depict the same flexibility as conversations with human counsellors \citep{ahmed2022generation}.

Given the high flexibility of large language models (LLMs) and their ability to generate human-like language, they are increasingly considered as tools to generate texts for sensitive use cases, such as mental health and counselling \citep{sharma2023cognitive, shen-etal-2020-counseling, ahmed2022generation, sharma2023human}. This bears several advantages, but also many safety hazards \citep{bender2021dangers}. Consequently, research on the topic is currently mainly limited to studying LLMs capabilities to perform specific behaviours in isolation – both of a longer conversation and of the user \citep{nori2023capabilities, shen-etal-2020-counseling, ahmed2022generation, ayers2023comparing}. In many such cases, existing datasets (i.e. \textsc{EmpatheticDialogues}) are used both to synthesise new conversational turns by the CA, and to evaluate the goodness of generations through similarity-based metrics or human empathy judgements \citep{li2021towards}.  
This system-oriented focus has led to a disregard of the user-side and a lack of resources and evaluations for studying real user interactions with LLM-based agents \citep{weidinger2023sociotechnical}. This is explored more frequently in the context of rule-based and retrieval-based conversational agents for counselling. For instance, \citet{he2022can} compared a rule-based MI-chatbot with a non-MI adherent conversational agent regarding user engagement, perceived empathy and therapeutic alliance. Given the new capabilities LLMs add to the picture of conversational AI, previous studies regarding user expectations for and behaviour towards CAs might not translate to these new technologies. Therefore, while researchers currently focus heavily on controlling specific LLM behaviours, one crucial aspect of CA-user interactions is overlooked: the unpredictability of user utterances. 

To address this, we present a dataset of user interactions about three target behaviours (healthy nutrition, less procrastination, increased sustainability) with two GPT-4-based conversational agents, where one system is prompted to adhere to MI principles, and the other is not. Each conversation spans 12 turns, of which 5 turns are generated by GPT-4. The beginning and end of each conversation are rule-based and reflect different phases of an MI-session. This dataset can drive the exploration of user behaviour when interacting with LLM-based conversational agents in the context of social influence dialogue systems and help shed light on the controllability of such systems within the context of multi-turn conversations. The insights gained in the analysis of the dataset have the potential to inform the user-oriented design of such systems in the future.

\section{Dataset}
The dataset stems from a preregistered online user study aimed at evaluating the controllability and efficacy of GPT-4-based motivational behaviour change support \footnote{Preregistering the study meant we publicly shared the study design, research questions, and analysis plan before data collection. Our time-stamped preregistration, along with extensive information about the study design, can be found \underline{\href{https://osf.io/yqe6w}{here}}.}. Overall, the dataset consists of 2149 conversational turns, collected in 185 chats with 164 study participants. The complete dataset can be found on \underline{\href{https://osf.io/c9nvm/}{OSF}}. In this section, we describe the study setup and the conversational system and provide information about the study participants and post-processing steps. 

\subsection{Study Setup}
At the beginning of the study, participants were provided with three personas to choose from, each focusing on a different target behaviour (healthier eating, sustainable living, less procrastination). These target behaviours were chosen, because they represent common nonmedical behaviour change goals \citep{ferrari2005prevalence, farhud2015impact, ONS2023, pinho2018exploring}. Participants were asked to choose the situation they identified with the most and put themselves in the persona's shoes during their interaction with the conversational system. This level of abstraction was added to avoid any potential harms interacting with the system might induce. However, in post-processing it became apparent that participants struggled to adhere to the level of abstraction, as most participants seemed to interact with the system from their own perspective.

Before the start of the conversation, participants were asked to what extent they identify with the chosen situation and are currently pursuing the target behaviour on two 10-point likert scales. 
Following \citet{he2022can}, we also collect six measures pertaining to participants' experiences conversing with the system and their perceived relationship with the conversational agent:
\begin{itemize}
    \item \textbf{Therapeutic Alliance.} The Working Alliance Inventory--short revised (WAI-SR) measures the relationship between a counsellor and a client \citep{wilmers2008deutschsprachige}. This questionnaire can be used to measure the degree to which interactions with the system are perceived as helpful and supportive by participants.
    \item \textbf{Perception of MI.} The Client Evaluation of Motivational Interviewing Scale (CEMI) is a measure designed to evaluate MI adherence of counsellors without the need of expert judgements. It is intended to be completed by clients directly after MI sessions, and serves as a proxy for the degree to which users feel that the spirit of MI was met during the interaction \citep{madson2013measuring, madson2015measuring, madson2016evaluating}.
    \item \textbf{User Engagement.} Since we focus on the conversational aspect of interactions, we follow \citep{he2022can} and use the User Engagement Scale--Short Form (UES-SF) without the aesthetic appeal subscale \citep{o2018practical, holdener2020applicability}.
    \item \textbf{Perceived Empathy \& Perceived Communication Competence.} We base the questions used for these measures on \citep{he2022can}. 
    \item \textbf{Readiness to Change.} Participants are asked to indicate their position on the \textit{Contemplation Ladder} \citep{biener1991contemplation, slavet2006marijuana}, a tool to measure readiness for behaviour change on an 11-point likert scale, both before and after interaction with the conversational system. The values indicated on the contemplation ladder can be translated into stages of change as defined by the transtheoretical model \citep{diclemente1998toward, velicer1998detailed}\footnote{the stages of change are precontemplation (no thoughts of changing), contemplation (beginning to consider change), preparation (preparing for change), action (taking steps to change), and maintenance (maintaining change).}.
\end{itemize}

\subsection{Conversational System}
Throughout the conversation, we used classifiers introduced and evaluated in previous work \citep{meyer2022glohbcd, meyer2023towards}, to evaluate participants' stance on behaviour change. Each user utterance was classified with a valence (change, sustain), a topic (Commitment, Taking Steps, Reason), and, if the topic is reason, a reason type (general, ability, need, desire). Each conversation consisted of three phases, based on the main MI processes and the structure of MI sessions \citep{miller2012motivational, clifford2016motivational}:

\textbf{Phase 1: Engaging and Focusing (4 turns)}.
The system introduces itself, asks the participant which behaviour they would like to change and why, asks a ``scaling question'' to determine their level of readiness to change and follows up on the scaling question by asking why the participant did not choose a higher or lower value. 

\textbf{Phase 2: Evoking (5 turns)}. During this part of the conversation, all bot utterances are generated by GPT-4. Participants were randomly assigned to one of two conditions:\newline
\textit{GPT-4}: The complete conversation history is passed to GPT-4 to solicit a bot utterance. \newline
\textit{MI-adapted GPT-4}: The classification of the current user utterance is used to identify a suitable bot behaviour, as previously mapped through MI-literature \citep{miller2012motivational, clifford2016motivational}. The complete conversation history is passed to GPT-4 with a system prompt setting the context of the conversation passing a definition of a suitable bot behaviour as given in MI-literature (see Table \ref{tab:adapted_prompt}). 

\begin{table*}[htbp]
    \centering
    \begin{tabular}{|p{\textwidth}|}
         \hline\textbf{System Prompt}: \\
         \textit{You are a counsellor and help the user with the goal [target behaviour]. 
        Never talk about yourself in the following, and concentrate fully on the user. Do not give active tips. If the user is talking about another topic, don't respond and lead them back to the [target behaviour]. Always speak in the second-person singular. You use the conversational strategy [action]. Definition of [action]: [description] Keep your answer as short as possible.}\\
        \textbf{Description for action Reframe:}\\
        A reframe is a reflection that highlights a different perspective in the client's statement. It is particularly useful for defusing sustain talk. A therapist can listen to sustain talk and reframe the statement into a neutral statement, change talk or an affirmation.\\\hline
    \end{tabular}
    \caption{System prompt passed to GPT-4 in the MI-adapted condition, with description of the sample action \textit{Reframe}}
    \label{tab:adapted_prompt}
\end{table*}

In both conditions, participants have to rate each bot utterance as either good/helpful, bad/unhelpful, or offensive/harmful and are given the option to explain their choice in free text format. We also set the same parameters for calls to the GPT-4 api in both conditions, setting the temperature to 0.7 and max\_tokens to 100.

\textbf{Phase 3: Conclusion (3 turns)}. The system summarises the conversation (generated by GPT) and invites the participant to add to the summary. The participant is then invited to define concrete next steps to take, and the system ends the conversation with a goodbye message. Except for the final bot utterance, participants are again prompted to rate the bot utterances and optionally explain their rating choice. 

\subsection{Participants}
The study was run on Prolific. It was performed in German with German native speakers. Each participant was paid €4 as compensation for their efforts, and the median completion time was 20.9 minutes. Participants were between 19 and 72 years old (mean=32.28, sd=9.45). 41.4\% were female. The majority of the participants were highly educated. 31.2\% had completed a university entry-level high school diploma and 56\% had at least a bachelor's degree. We provide demographic data for each participant in the dataset. 

Most participants chose procrastination as their area to work on, with the least participants interested in increasing sustainability (see Table \ref{tab:distr}). Participants generally had a high level of identification with their chosen target behaviour (only 25\% indicated a level of identification of 7 or less). Mean readiness to change at the beginning of the interaction was 6.25 (sd=2.27, min=0, max=10). This indicates that the interventions tested were relevant to the chosen participant pool and that the conversations collected can be expected to reflect realistic user-chatbot interactions. To ensure the validity of the post-test questionnaires, participants had to answer an attention check. 

\begin{table}[htbp]
\centering
\begin{tabular}{lcccc}
& healthier eating & sustainable living & less procrastination & sum \\\toprule
MI-adapted GPT-4 & 26 & 2 & 51 & 79 \\\hline
GPT-4 & 21 & 10 & 46 & 78 \\\bottomrule
\end{tabular}
\caption{Number of chats per condition and target behaviour.}
\label{tab:distr}
\end{table}

\subsection{Post-Processing}
We share the raw collected data in its original state. In addition, we apply a number of post-processing steps and share the processed data in separate files. In the following, we summarise all post-processing steps applied to the dataset.

5 participants did not pass the attention check, and 16 participants led multiple (at least two) chats with the system. Since the chat interactions of these participants can nevertheless give valuable information, we decided to keep the data points in the dataset, even though the measures collected might not be valid in those cases. We provide a separate file with duplicate submission IDs and mark participants who did not pass the attention checks. 16 chats experienced technical issues. For example, these led to users having to send a message more than once before the system replied. Again, we marked these instances, but kept the associated data, as it can potentially be used to evaluate how users handle such technical errors, especially when the underlying conversational agent is highly capable as is the case for GPT-4. 

We also calculate the overall results for each post-processing questionnaire (plus the values for each user engagement subscale), the $\Delta readiness to change$ by subtracting pre-conversation from post-conversation values, and translate the readiness to change indicated by participants at the start of the conversation to the corresponding stage of change as defined in \citep{slavet2006marijuana, biener1991contemplation}. 

Since we notice that in many instances, user utterances are not directly related to change, instead constituting reactions to bot utterances, we run a prefilter\footnote{\url{https://huggingface.co/selmey/behaviour_change_prefilter_german}} trained on the same dataset as the valence, topic, and reason type classifiers over each user utterance. This prefilter classifies an utterance as change related or Follow/Neutral (not directly related to behaviour change). We then introduce follow/neutral as a third label for user utterance valences, resulting in valence distinctions of change talk, sustain talk, or follow/neutral. 

Finally, to make the dataset appealing to a wider audience, we translate all conversational turns and rating explanations into English using a pre-trained machine translation model\footnote{\url{https://huggingface.co/Helsinki-NLP/opus-mt-de-en} -- While some basic quality control of the translations was conducted, subtle nuances in language may not be fully captured. Readers should be aware of potential translation imperfections.} \citep{TiedemannThottingal:EAMT2020}. We share both the German original and the English translations of the conversational data. 
\section{Potential Applications}
Due to the various variables explored in the dataset, its potential applications are vast, and range from behavioural analysis to the training of classifiers and generative models to improve LLM-outputs in the context. Thus, the dataset is relevant for research in human-computer interaction and information behaviour, as well as natural language processing. Below, we share two potential applications of the dataset, as well as a non-exhaustive list of research questions to be explored. Further relevant research questions related to the data set are described in our previous work \citep{meyer2022m, meyer2021natural} and in the preregistration of the user study \citep{meyer2023evaluating}.

\subsection{Exploration of User Expectations}
The first message sent to participants in each conversation informed them that the goal of the conversation is not to give advice, but instead to help them reflect about behaviour change, in order to create their own plans for change. Although this message told participants not to expect advice and facts, a non-negligible share of participants voiced discontent with the lack of advice in the MI-adapted condition, either in their rating justifications, or in the chat itself, as can be seen below:
\begin{quote}
    \textbf{CA:} Understood! Finally, please tell me again what your next step will be. \\
         \textbf{Participant:} You tell me. 
\end{quote}

In contrast, other users took the questions posed by the conversational system as nudges to reflect and come up with solutions themselves. A qualitative analysis of the chats could shed light on the expectations different users pose to such systems. Many of the variables we collected (i.e. demographic data, readiness to change, chosen target behaviour) have the potential to influence user behaviour and can be explored as differentiating aspects. 

\subsection{Anticipating Information Needs}
Throughout the conversations, the users shared information needs, both explicitly and implicitly. Note the following example:

\begin{quote}
    \textbf{CA:} [...] How will you make sure you follow your diet plan? \\
         \textbf{Participant:} That's the hard part, it has to be simple, not complicated and expensive food, and it has to be tasty too. I can never be sure, I have to try it out.
\end{quote}

Even though the participant does not directly ask the conversational agent for help here, they still voice an information need that could be addressed by the system. Being able to anticipate these implicit information needs has the potential to greatly improve conversational search. This dataset can serve as a resource to analyse how users implicitly voice information needs in conversations with LLMs. Future research could then combine LLMs to provide flexible emotional support with retrieval to resolve implicitly expressed information needs identified through interactions with the LLM.  

\subsection{Relevant Research Questions}
\textbf{How does user interaction with the system differ based on conversational condition?}\newline
Variables to explore include utterance classifications, bot utterance ratings given by users, self-disclosure by the user, and how cooperative the user is in interaction with the system. 

\textbf{Can conversations with a chatbot be used to draw conclusions about a user's readiness to change their behaviour?}\newline
Another avenue to explore would be whether the conversational logs can be used to draw conclusions about chat success.

\textbf{What impact does user behaviour have on chat success?}\newline
Research on early LLM-based generative chatbots in the context of general chit-chat has shown that unclear user utterances have a significant impact in the ability of such systems to perform well \citep{see2021understanding}. Other user behaviour aspects could have similar effects.

\textbf{Which bot behaviours are most popular with users, and which bot behaviours are most effective in inducing an increase in readiness to change?} \newline
In this context, it is also worth exploring to what degree those groups intersect. 

\textbf{How do users interact with LLM-based systems in the context of behaviour change, compared to more restrictive (rule- or retrieval-based) systems? }\newline
The dataset can be used to expand on existing literature on user interactions with those traditional conversational agents, i.e. regarding user perceptions of chatbot mistakes \citep{de2023we} and self-disclosure in conversation with different types of chatbots \citep{kang2023counseling, lee2020hear, lee2022influence}.
\section{Ethical Considerations}
Participants were informed about the fact they are conversing with a GPT-based conversational agent in the informed consent. They were also explicitly told that their utterances are passed on to a third party, and advised not to share any information they would not feel comfortable disclosing in an anonymous forum. We urged participants to take on the role of one of three personas. Although we observed that the majority of participants did not consistently maintain these personas during their interactions with the system, they were repeatedly reminded of the experimental nature of the system through the solicitation of a judgement about response helpfulness and safety after each bot turn.
\section{Conclusion}
In this resource paper, we describe the collection process and potential applications of a dataset of interactions between 164 users and two GPT-4 based conversational systems. In addition to the conversations between users and systems, the dataset contains a variety of information that can be used to analyse user behaviour and conversation success: pre- and post-test measure results, utterance classifications regarding user's stance on behaviour change, ratings of LLM-generated bot-turns and rating explanations. Interactions with the MI-adapted condition also include information on the type of utterance prompted to GPT-4. The dataset can be used to analyse user and GPT-4 behaviour in the context of behaviour change, and serves as a valuable resource for the improvement of such systems in the future. 

\bibliographystyle{unsrtnat}
\bibliography{sample-base} 

\begin{thebibliography}{54}
\providecommand{\natexlab}[1]{#1}
\providecommand{\url}[1]{\texttt{#1}}
\expandafter\ifx\csname urlstyle\endcsname\relax
  \providecommand{\doi}[1]{doi: #1}\else
  \providecommand{\doi}{doi: \begingroup \urlstyle{rm}\Url}\fi

\bibitem[Elsholz et~al.(2019)Elsholz, Chamberlain, and Kruschwitz]{elsholz2019exploring}
Ela Elsholz, Jon Chamberlain, and Udo Kruschwitz.
\newblock {Exploring language style in chatbots to increase perceived product value and user engagement}.
\newblock In \emph{Proceedings of the 2019 Conference on Human Information Interaction and Retrieval}, pages 301--305, 2019.

\bibitem[Papenmeier et~al.(2021)Papenmeier, Kern, Hienert, Sliwa, Aker, and Fuhr]{papenmeier2021dataset}
Andrea Papenmeier, Dagmar Kern, Daniel Hienert, Alfred Sliwa, Ahmet Aker, and Norbert Fuhr.
\newblock {Dataset of Natural Language Queries for E-Commerce}.
\newblock In \emph{Proceedings of the 2021 Conference on Human Information Interaction and Retrieval}, pages 307--311, 2021.

\bibitem[Papenmeier et~al.(2022)Papenmeier, Frummet, and Kern]{papenmeier2022mhm}
Andrea Papenmeier, Alexander Frummet, and Dagmar Kern.
\newblock {“Mhm...”--Conversational Strategies For Product Search Assistants}.
\newblock In \emph{Proceedings of the 2022 Conference on Human Information Interaction and Retrieval}, pages 36--46, 2022.

\bibitem[Barko-Sherif et~al.(2020)Barko-Sherif, Elsweiler, and Harvey]{barko2020conversational}
Sabrina Barko-Sherif, David Elsweiler, and Morgan Harvey.
\newblock {Conversational agents for recipe recommendation}.
\newblock In \emph{Proceedings of the 2020 Conference on Human Information Interaction and Retrieval}, pages 73--82, 2020.

\bibitem[Frummet et~al.(2022)Frummet, Elsweiler, and Ludwig]{frummet2022can}
Alexander Frummet, David Elsweiler, and Bernd Ludwig.
\newblock {“What Can I Cook with these Ingredients?”-Understanding Cooking-Related Information Needs in Conversational Search}.
\newblock \emph{ACM Transactions on Information Systems (TOIS)}, 40\penalty0 (4):\penalty0 1--32, 2022.

\bibitem[Nouri et~al.(2020)Nouri, Sim, Fourney, and White]{nouri2020step}
Elnaz Nouri, Robert Sim, Adam Fourney, and Ryen~W White.
\newblock {Step-wise recommendation for complex task support}.
\newblock In \emph{Proceedings of the 2020 Conference on Human Information Interaction and Retrieval}, pages 203--212, 2020.

\bibitem[Elsweiler et~al.(2011)Elsweiler, Wilson, and Lunn]{elsweiler2011understanding}
David Elsweiler, Max~L Wilson, and Brian~Kirkegaard Lunn.
\newblock {Understanding casual-leisure information behaviour}.
\newblock In \emph{New directions in information behaviour}, volume~1, pages 211--241. Emerald Group Publishing Limited, 2011.

\bibitem[Ruthven(2019)]{ruthven2019making}
Ian Ruthven.
\newblock {Making meaning: A focus for information interactions research}.
\newblock In \emph{Proceedings of the 2019 conference on human information interaction and retrieval}, pages 163--171, 2019.

\bibitem[Lopatovska and Davis(2023)]{lopatovska2023designing}
Irene Lopatovska and Jessika Davis.
\newblock {Designing Supportive Conversational Agents With and For Teens}.
\newblock In \emph{Proceedings of the 2023 Conference on Human Information Interaction and Retrieval}, pages 328--332, 2023.

\bibitem[Rapp et~al.(2021)Rapp, Curti, and Boldi]{rapp2021human}
Amon Rapp, Lorenzo Curti, and Arianna Boldi.
\newblock {The human side of human-chatbot interaction: A systematic literature review of ten years of research on text-based chatbots}.
\newblock \emph{International Journal of Human-Computer Studies}, 151:\penalty0 102630, 2021.

\bibitem[Meyer(2022)]{meyer2022m}
Selina Meyer.
\newblock {“I’m at my wits’ end”-Anticipating Information Needs and Appropriate Support Strategies in Behaviour Change}.
\newblock In \emph{Proceedings of the 2022 Conference on Human Information Interaction and Retrieval}, pages 396--399, 2022.

\bibitem[Chawla et~al.(2023)Chawla, Shi, Zhang, Lucas, Yu, and Gratch]{chawla-etal-2023-social}
Kushal Chawla, Weiyan Shi, Jingwen Zhang, Gale Lucas, Zhou Yu, and Jonathan Gratch.
\newblock {Social Influence Dialogue Systems: A Survey of Datasets and Models For Social Influence Tasks}.
\newblock In \emph{Proceedings of the 17th Conference of the European Chapter of the Association for Computational Linguistics}, pages 750--766, Dubrovnik, Croatia, May 2023. Association for Computational Linguistics.
\newblock URL \url{https://aclanthology.org/2023.eacl-main.53}.

\bibitem[Miller and Rollnick(2012)]{miller2012motivational}
William~R Miller and Stephen Rollnick.
\newblock \emph{{Motivational interviewing: Helping people change}}.
\newblock Guilford press, 2012.

\bibitem[Xu and Zhuang(2022)]{xu2022survey}
Bei Xu and Ziyuan Zhuang.
\newblock Survey on psychotherapy chatbots.
\newblock \emph{Concurrency and Computation: Practice and Experience}, 34\penalty0 (7):\penalty0 e6170, 2022.

\bibitem[He et~al.(2022)He, Basar, Wiers, Antheunis, and Krahmer]{he2022can}
Linwei He, Erkan Basar, Reinout~W Wiers, Marjolijn~L Antheunis, and Emiel Krahmer.
\newblock {Can chatbots help to motivate smoking cessation? A study on the effectiveness of motivational interviewing on engagement and therapeutic alliance}.
\newblock \emph{BMC Public Health}, 22\penalty0 (1):\penalty0 726, 2022.

\bibitem[Park et~al.(2019)Park, Choi, Lee, Oh, Kim, La, Lee, Suh, et~al.]{park2019designing}
SoHyun Park, Jeewon Choi, Sungwoo Lee, Changhoon Oh, Changdai Kim, Soohyun La, Joonhwan Lee, Bongwon Suh, et~al.
\newblock {Designing a chatbot for a brief motivational interview on stress management: Qualitative case study}.
\newblock \emph{Journal of medical Internet research}, 21\penalty0 (4):\penalty0 e12231, 2019.

\bibitem[Schulman et~al.(2011)Schulman, Bickmore, and Sidner]{schulman2011intelligent}
Daniel Schulman, Timothy~W Bickmore, and Candace~L Sidner.
\newblock {An Intelligent Conversational Agent for Promoting Long-Term Health Behavior Change Using Motivational Interviewing.}
\newblock In \emph{AAAI Spring Symposium: AI and Health Communication}, pages 61--64, 2011.

\bibitem[Olafsson et~al.(2019)Olafsson, O'Leary, and Bickmore]{olafsson2019coerced}
Stefan Olafsson, Teresa O'Leary, and Timothy Bickmore.
\newblock {Coerced change-talk with conversational agents promotes confidence in behavior change}.
\newblock In \emph{Proceedings of the 13th EAI International Conference on Pervasive Computing Technologies for Healthcare}, pages 31--40, 2019.

\bibitem[Boustani et~al.(2021)Boustani, Lunn, Visser, Lisetti, et~al.]{boustani2021development}
Maya Boustani, Stephanie Lunn, Ubbo Visser, Christine Lisetti, et~al.
\newblock {Development, Feasibility, Acceptability, and Utility of an Expressive Speech-Enabled Digital Health Agent to Deliver Online, Brief Motivational Interviewing for Alcohol Misuse: Descriptive Study}.
\newblock \emph{Journal of medical Internet research}, 23\penalty0 (9):\penalty0 e25837, 2021.

\bibitem[Samrose and Hoque(2022)]{samrose2022mia}
Samiha Samrose and Ehsan Hoque.
\newblock {MIA: Motivational Interviewing Agent for Improving Conversational Skills in Remote Group Discussions}.
\newblock \emph{Proceedings of the ACM on Human-Computer Interaction}, 6\penalty0 (GROUP):\penalty0 1--24, 2022.

\bibitem[Ahmed et~al.(2022)Ahmed, Keilty, Cooper, Selby, and Rose]{ahmed2022generation}
Imtihan Ahmed, Eric Keilty, Carolynne Cooper, Peter Selby, and Jonathan Rose.
\newblock {Generation and Classification of Motivational-Interviewing-Style Reflections for Smoking Behaviour Change Using Few-Shot Learning with Transformers}.
\newblock 2022.

\bibitem[Sharma et~al.(2023{\natexlab{a}})Sharma, Rushton, Lin, Wadden, Lucas, Miner, Nguyen, and Althoff]{sharma2023cognitive}
Ashish Sharma, Kevin Rushton, Inna~Wanyin Lin, David Wadden, Khendra~G Lucas, Adam~S Miner, Theresa Nguyen, and Tim Althoff.
\newblock {Cognitive Reframing of Negative Thoughts through Human-Language Model Interaction}.
\newblock \emph{arXiv preprint arXiv:2305.02466}, 2023{\natexlab{a}}.

\bibitem[Shen et~al.(2020)Shen, Welch, Mihalcea, and P{\'e}rez-Rosas]{shen-etal-2020-counseling}
Siqi Shen, Charles Welch, Rada Mihalcea, and Ver{\'o}nica P{\'e}rez-Rosas.
\newblock {Counseling-Style Reflection Generation Using Generative Pretrained Transformers with Augmented Context}.
\newblock In \emph{Proceedings of the 21th Annual Meeting of the Special Interest Group on Discourse and Dialogue}, pages 10--20, 1st virtual meeting, July 2020. Association for Computational Linguistics.
\newblock URL \url{https://aclanthology.org/2020.sigdial-1.2}.

\bibitem[Sharma et~al.(2023{\natexlab{b}})Sharma, Lin, Miner, Atkins, and Althoff]{sharma2023human}
Ashish Sharma, Inna~W Lin, Adam~S Miner, David~C Atkins, and Tim Althoff.
\newblock {Human--AI collaboration enables more empathic conversations in text-based peer-to-peer mental health support}.
\newblock \emph{Nature Machine Intelligence}, 5\penalty0 (1):\penalty0 46--57, 2023{\natexlab{b}}.

\bibitem[Bender et~al.(2021)Bender, Gebru, McMillan-Major, and Shmitchell]{bender2021dangers}
Emily~M. Bender, Timnit Gebru, Angelina McMillan-Major, and Shmargaret Shmitchell.
\newblock {On the Dangers of Stochastic Parrots: Can Language Models Be Too Big?}
\newblock In \emph{Proceedings of the 2021 ACM Conference on Fairness, Accountability, and Transparency}, FAccT '21, page 610–623, New York, NY, USA, 2021. Association for Computing Machinery.
\newblock ISBN 9781450383097.
\newblock \doi{10.1145/3442188.3445922}.
\newblock URL \url{https://doi.org/10.1145/3442188.3445922}.

\bibitem[Nori et~al.(2023)Nori, King, McKinney, Carignan, and Horvitz]{nori2023capabilities}
Harsha Nori, Nicholas King, Scott~Mayer McKinney, Dean Carignan, and Eric Horvitz.
\newblock {Capabilities of gpt-4 on medical challenge problems}.
\newblock \emph{arXiv preprint arXiv:2303.13375}, 2023.

\bibitem[Ayers et~al.(2023)Ayers, Poliak, Dredze, Leas, Zhu, Kelley, Faix, Goodman, Longhurst, Hogarth, and Smith]{ayers2023comparing}
John~W Ayers, Adam Poliak, Mark Dredze, Eric~C Leas, Zechariah Zhu, Jessica~B Kelley, Dennis~J Faix, Aaron~M Goodman, Christopher~A Longhurst, Michael Hogarth, and Davey~M Smith.
\newblock {Comparing Physician and Artificial Intelligence Chatbot Responses to Patient Questions Posted to a Public Social Media Forum}.
\newblock \emph{JAMA internal medicine}, 183\penalty0 (6):\penalty0 589—596, June 2023.
\newblock ISSN 2168-6106.
\newblock \doi{10.1001/jamainternmed.2023.1838}.
\newblock URL \url{https://jamanetwork.com/journals/jamainternalmedicine/articlepdf/2804309/jamainternal_ayers_2023_oi_230030_1681999216.70842.pdf}.

\bibitem[Li et~al.(2021)Li, Li, Ning, Xia, Guo, Wei, Cui, and Wang]{li2021towards}
Yanran Li, Ke~Li, Hongke Ning, Xiaoqiang Xia, Yalong Guo, Chen Wei, Jianwei Cui, and Bin Wang.
\newblock {Towards an online empathetic chatbot with emotion causes}.
\newblock In \emph{Proceedings of the 44th International ACM SIGIR Conference on Research and Development in Information Retrieval}, pages 2041--2045, 2021.

\bibitem[Weidinger et~al.(2023)Weidinger, Rauh, Marchal, Manzini, Hendricks, Mateos-Garcia, Bergman, Kay, Griffin, Bariach, Gabriel, Rieser, and Isaac]{weidinger2023sociotechnical}
Laura Weidinger, Maribeth Rauh, Nahema Marchal, Arianna Manzini, Lisa~Anne Hendricks, Juan Mateos-Garcia, Stevie Bergman, Jackie Kay, Conor Griffin, Ben Bariach, Iason Gabriel, Verena Rieser, and William Isaac.
\newblock {Sociotechnical Safety Evaluation of Generative AI Systems}, 2023.

\bibitem[Ferrari et~al.(2005)Ferrari, O'Callaghan, and Newbegin]{ferrari2005prevalence}
Joseph~R Ferrari, Jean O'Callaghan, and Ian Newbegin.
\newblock {Prevalence of procrastination in the United States, United Kingdom, and Australia: arousal and avoidance delays among adults.}
\newblock \emph{North American Journal of Psychology}, 7\penalty0 (1), 2005.

\bibitem[Farhud(2015)]{farhud2015impact}
Dariush~D Farhud.
\newblock {Impact of lifestyle on health}.
\newblock \emph{Iranian journal of public health}, 44\penalty0 (11):\penalty0 1442, 2015.

\bibitem[{Office for National Statistics}()]{ONS2023}
{Office for National Statistics}.
\newblock {Most adults report making some changes to their lifestyle for environmental reasons}.
\newblock URL \url{https://www.ons.gov.uk/peoplepopulationandcommunity/wellbeing/articles/mostadultsreportmakingsomechangestotheirlifestyleforenvironmentalreasons/2023-07-05}.

\bibitem[Pinho et~al.(2018)Pinho, Mackenbach, Charreire, Oppert, B{\'a}rdos, Glonti, Rutter, Compernolle, De~Bourdeaudhuij, Beulens, et~al.]{pinho2018exploring}
MGM Pinho, JD~Mackenbach, H{\'e}l{\`e}ne Charreire, J-M Oppert, H~B{\'a}rdos, K~Glonti, H~Rutter, Sofie Compernolle, Ilse De~Bourdeaudhuij, JWJ Beulens, et~al.
\newblock {Exploring the relationship between perceived barriers to healthy eating and dietary behaviours in European adults}.
\newblock \emph{European journal of nutrition}, 57:\penalty0 1761--1770, 2018.

\bibitem[Wilmers et~al.(2008)Wilmers, Munder, Leonhart, Herzog, Plassmann, Barth, and Linster]{wilmers2008deutschsprachige}
Fabian Wilmers, Thomas Munder, Rainer Leonhart, Thomas Herzog, Reinhard Plassmann, J{\"u}rgen Barth, and Hans~Wolfgang Linster.
\newblock {Die deutschsprachige Version des Working Alliance Inventory-short revised (WAI-SR)-Ein schulen{\"u}bergreifendes, {\"o}konomisches und empirisch validiertes Instrument zur Erfassung der therapeutischen Allianz}.
\newblock \emph{Klinische Diagnostik und Evaluation}, 1\penalty0 (3):\penalty0 343--358, 2008.

\bibitem[Madson et~al.(2013)Madson, Mohn, Zuckoff, Schumacher, Kogan, Hutchison, Magee, and Stein]{madson2013measuring}
Michael~B Madson, Richard~S Mohn, Allan Zuckoff, Julie~A Schumacher, Jane Kogan, Shari Hutchison, Emily Magee, and Bradley Stein.
\newblock {Measuring client perceptions of motivational interviewing: factor analysis of the Client Evaluation of Motivational Interviewing scale}.
\newblock \emph{Journal of Substance Abuse Treatment}, 44\penalty0 (3):\penalty0 330--335, 2013.

\bibitem[Madson et~al.(2015)Madson, Mohn, Schumacher, and Landry]{madson2015measuring}
Michael~B Madson, Richard~S Mohn, Julie~A Schumacher, and Alicia~S Landry.
\newblock {Measuring client experiences of motivational interviewing during a lifestyle intervention}.
\newblock \emph{Measurement and Evaluation in Counseling and Development}, 48\penalty0 (2):\penalty0 140--151, 2015.

\bibitem[Madson et~al.(2016)Madson, Villarosa, Schumacher, and Mohn]{madson2016evaluating}
Michael~B Madson, Margo~C Villarosa, Julie~A Schumacher, and Richard~S Mohn.
\newblock {Evaluating the validity of the client evaluation of motivational interviewing scale in a brief motivational intervention for college student drinkers}.
\newblock \emph{Journal of substance abuse treatment}, 65:\penalty0 51--57, 2016.

\bibitem[O’Brien et~al.(2018)O’Brien, Cairns, and Hall]{o2018practical}
Heather~L O’Brien, Paul Cairns, and Mark Hall.
\newblock {A practical approach to measuring user engagement with the refined user engagement scale (UES) and new UES short form}.
\newblock \emph{International Journal of Human-Computer Studies}, 112:\penalty0 28--39, 2018.

\bibitem[Holdener et~al.(2020)Holdener, Gut, Angerer, et~al.]{holdener2020applicability}
Marianne Holdener, Alain Gut, Alfred Angerer, et~al.
\newblock {Applicability of the user engagement scale to mobile health: a survey-based quantitative study}.
\newblock \emph{JMIR mHealth and uHealth}, 8\penalty0 (1):\penalty0 e13244, 2020.

\bibitem[Biener and Abrams(1991)]{biener1991contemplation}
Lois Biener and David~B Abrams.
\newblock {The Contemplation Ladder: validation of a measure of readiness to consider smoking cessation.}
\newblock \emph{Health psychology}, 10\penalty0 (5):\penalty0 360, 1991.

\bibitem[Slavet et~al.(2006)Slavet, Stein, Colby, Barnett, Monti, Golembeske~Jr, and Lebeau-Craven]{slavet2006marijuana}
James~D Slavet, LAR Stein, Suzanne~M Colby, Nancy~P Barnett, Peter~M Monti, Charles Golembeske~Jr, and Rebecca Lebeau-Craven.
\newblock {The Marijuana Ladder: Measuring motivation to change marijuana use in incarcerated adolescents}.
\newblock \emph{Drug and Alcohol Dependence}, 83\penalty0 (1):\penalty0 42--48, 2006.

\bibitem[DiClemente and Prochaska(1998)]{diclemente1998toward}
Carlo~C DiClemente and James~O Prochaska.
\newblock {Toward a comprehensive, transtheoretical model of change: Stages of change and addictive behaviors.}
\newblock 1998.

\bibitem[Velicer et~al.(1998)Velicer, Prochaska, Fava, Norman, and Redding]{velicer1998detailed}
WF~Velicer, JO~Prochaska, JL~Fava, GJ~Norman, and CA~Redding.
\newblock {Detailed overview of the transtheoretical model}.
\newblock \emph{Homeostasis}, 38:\penalty0 216--33, 1998.

\bibitem[Meyer and Elsweiler(2022)]{meyer2022glohbcd}
Selina Meyer and David Elsweiler.
\newblock {GLoHBCD: A Naturalistic German Dataset for Language of Health Behaviour Change on Online Support Forums}.
\newblock In \emph{Proceedings of the Thirteenth Language Resources and Evaluation Conference}, pages 2226--2235, 2022.

\bibitem[Meyer and Elsweiler(2023{\natexlab{a}})]{meyer2023towards}
Selina Meyer and David Elsweiler.
\newblock Towards cross-content conversational agents for behaviour change: Investigating domain independence and the role of lexical features in written language around change.
\newblock In \emph{Proceedings of the 5th International Conference on Conversational User Interfaces}, pages 1--13, 2023{\natexlab{a}}.

\bibitem[Clifford and Curtis(2016)]{clifford2016motivational}
Dawn Clifford and Laura Curtis.
\newblock \emph{Motivational interviewing in nutrition and fitness}.
\newblock Guilford Publications, 2016.

\bibitem[Tiedemann and Thottingal(2020)]{TiedemannThottingal:EAMT2020}
J{\"o}rg Tiedemann and Santhosh Thottingal.
\newblock {OPUS-MT} — {B}uilding open translation services for the {W}orld.
\newblock In \emph{Proceedings of the 22nd Annual Conferenec of the European Association for Machine Translation (EAMT)}, Lisbon, Portugal, 2020.

\bibitem[Meyer(2021)]{meyer2021natural}
Selina Meyer.
\newblock {Natural Language Stage of Change Modelling for “Motivationally-driven” Weight Loss Support}.
\newblock In \emph{Proceedings of the 2021 International Conference on Multimodal Interaction}, pages 807--811, 2021.

\bibitem[Meyer and Elsweiler(2023{\natexlab{b}})]{meyer2023evaluating}
Selina Meyer and David Elsweiler.
\newblock {Evaluating the Efficacy, Controllability, and Safety of LLM-driven Conversational Agents to Support Behaviour Change}.
\newblock 2023{\natexlab{b}}.

\bibitem[See and Manning(2021)]{see2021understanding}
Abigail See and Christopher~D Manning.
\newblock {Understanding and predicting user dissatisfaction in a neural generative chatbot}.
\newblock In \emph{Proceedings of the 22nd Annual Meeting of the Special Interest Group on Discourse and Dialogue}, pages 1--12, 2021.

\bibitem[de~S{\'a}~Siqueira et~al.(2023)de~S{\'a}~Siqueira, M{\"u}ller, and Bosse]{de2023we}
Marianna~A de~S{\'a}~Siqueira, Barbara~CN M{\"u}ller, and Tibor Bosse.
\newblock {When do we accept mistakes from chatbots? The impact of human-like communication on user experience in chatbots that make mistakes}.
\newblock \emph{International Journal of Human--Computer Interaction}, pages 1--11, 2023.

\bibitem[Kang and Kang(2023)]{kang2023counseling}
Eunbin Kang and Youn~Ah Kang.
\newblock {Counseling chatbot design: The effect of anthropomorphic chatbot characteristics on user self-disclosure and companionship}.
\newblock \emph{International Journal of Human--Computer Interaction}, pages 1--15, 2023.

\bibitem[Lee et~al.(2020)Lee, Yamashita, Huang, and Fu]{lee2020hear}
Yi-Chieh Lee, Naomi Yamashita, Yun Huang, and Wai Fu.
\newblock {" I hear you, I feel you": encouraging deep self-disclosure through a chatbot}.
\newblock In \emph{Proceedings of the 2020 CHI conference on human factors in computing systems}, pages 1--12, 2020.

\bibitem[Lee et~al.(2022)Lee, Lee, and Lee]{lee2022influence}
Jieon Lee, Daeho Lee, and Jae-gil Lee.
\newblock {Influence of Rapport and Social Presence with an AI Psychotherapy Chatbot on Users’ Self-Disclosure}.
\newblock \emph{International Journal of Human--Computer Interaction}, pages 1--12, 2022.

\end{thebibliography}

\end{document}